\documentclass[11pt, oneside]{article}

\usepackage[a4paper, total={6.2in, 9in}]{geometry}
\pdfoutput=1
\usepackage{graphicx}
\usepackage{amssymb}
\usepackage[table]{xcolor}
\usepackage{xcolor}
\usepackage[T1]{fontenc}
\usepackage{lmodern}
\usepackage[bottom]{footmisc}
\usepackage[normalem]{ulem} 

\usepackage{multirow,booktabs,array,longtable,caption}

\PassOptionsToPackage{hyphens}{url}\usepackage[hidelinks]{hyperref}
\usepackage{fancyhdr}                           
\fancypagestyle{firstpage} 
{
   \fancyhf{}
   \fancyfoot[C]{\vspace{-12mm}\thepage}

   \fancyfoot[L]{\vspace{-5mm}\noindent\rule{6.2in}{0.4pt}
     \scriptsize{
       \textcopyright \hspace{0.5mm} Barclays Bank PLC 2020 \\
       This work is licensed under a \href{https://creativecommons.org/licenses/by/4.0/}
       {Creative Commons Attribution 4.0 International License (CC BY).}  \\ 
       Provided you adhere to the CC BY license, including as to attribution, you are
       free to copy and redistribute this work in any medium or format and remix, transform,
       and build upon the work for any purpose, even commercially.                  \\
       BARCLAYS is a registered trade mark of Barclays Bank PLC, all rights are reserved. \\
     }
   }
}

\title{\vspace{-2.5cm} Historical Context and Key Features of \\ Digital Money Tokens}

\usepackage{anyfontsize}
\author{%
  \hspace{-1.5em}
  \begin{tabular}{c} 
    {\fontsize{10.75}{1cm}\selectfont Shreepad Shukla} \\ 
    {\fontsize{10.75}{1cm}\selectfont Chief Technology Office} \\
    {\fontsize{10.75}{1cm}\selectfont Barclays} \\
  \end{tabular} 
  \hspace{-1.5em}
}

\date{August 25, 2020}

\begin{document}
\maketitle
\thispagestyle{firstpage} 
\vspace{-1cm}
\begin{abstract}
  \noindent
    Digital money tokens have attracted the attention of financial
    institutions, central banks, regulators, international associations and fintechs.
    Their research and experimentation with digital money tokens has included creating
    innovative technical and operational frameworks. 
    In this paper, we present a `money tree' which places this recent concept
    of digital money tokens into a historical context by illustrating their evolution from 
    more traditional forms of money. 
    We then identify key features of digital money tokens with options and
    examples.
    We hope this paper will be of interest to the financial services
    industry and we look forward to feedback.
\end{abstract}

\vspace{5mm}

\section{Introduction}
\label{sec:introduction}

Research and experimentation with digital money tokens is being conducted by 
financial institutions \cite{jpmcoin}, 
central banks \cite{ekrona, mas-ubin-5}, 
regulators \cite{ecconsultation, swiss-finma-newguidelines}, 
international associations \cite{fsb-stablecoin-consultation, iosco-stablecoin-report} 
and fintechs \cite{fnality, libra}.
As an emerging technology, the innovative technical and operational frameworks are often
bespoke.
The terminology used to describe forms of digital money
tokens is also inconsistent across these frameworks.

It can therefore be difficult to identify and understand the key features that distinguish
different forms of digital money tokens, creating challenges in specifying and
communicating new products and also potentially creating inappropriate opportunities for
regulatory arbitrage across jurisdictions.
To address this, some industry bodies are exploring standard taxonomies for 
digital money tokens.
For example: 
(i) the Global Financial Markets Association (GFMA) has proposed an initial starting point
for a classification of crypto-assets in their response \cite{gfma-bcbs-response}
to a Basel Committee on Banking Supervision (BCBS) consultation
on the treatment of crypto-assets \cite{bcbs-cryptoasset-consultation} and also in
their response \cite{gfma-fsb-response} to an FSB
consultation on challenges raised by global stablecoins \cite{fsb-stablecoin-consultation}
and
(ii) the Association for Financial Markets in Europe (AFME) has recommended the
publication of a more detailed taxonomy on the classification of
crypto-assets in their response \cite{afme-ec-response} to the European Commission 
consultation on a framework for crypto-asset markets \cite{ecconsultation}.

This paper provides a historical context for digital money tokens
by placing them within a `money tree' illustrating their evolution from more
traditional forms of money.
We then identify key features of digital money tokens with options and
examples.

We hope this money tree and list of key features will be of
interest to the financial services industry. 
We look forward to feedback and continuing industry collaboration on the 
classification of digital money tokens.

\pagebreak

\section{The Money Tree}
\label{sec:evolution-money}

Money is one of the fundamental inventions that support human societies and as a result its
history has been studied extensively \cite{mitchell-innes-money-paper,
uot-moneyevolution-paper, moneyevolution-book, wealthnations-book,
banks-ages-book}, from the first forms of account-based money and commodity money,
through the development of fiat money and on to new digital forms of money.

\begin{figure}[h!]
\begin{center}
\includegraphics[height=0.45\paperheight]{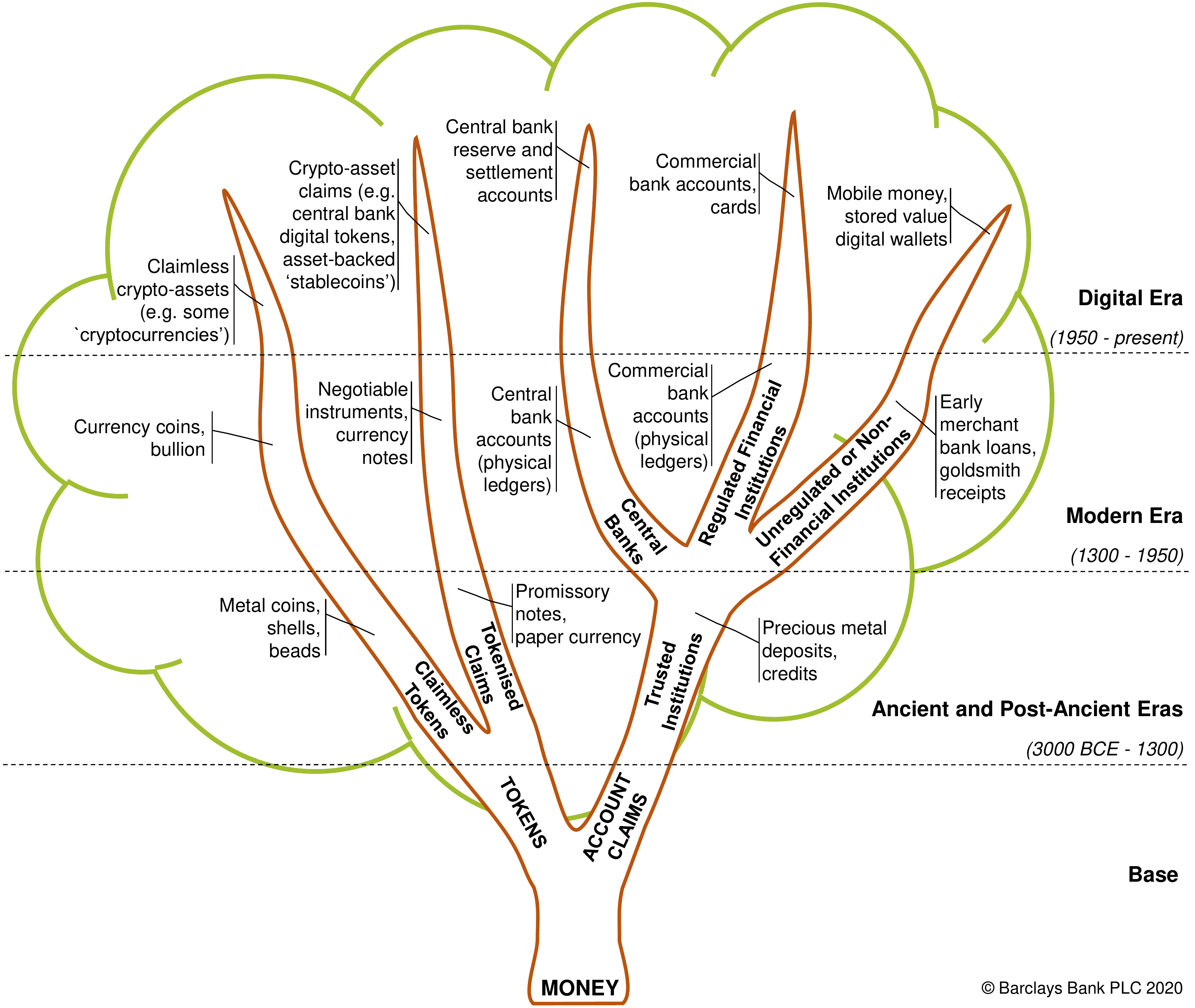}

\vspace{5mm}

\parbox{15cm}{\caption{\footnotesize {The `money tree' depicts the evolution of forms
of money from the ancient era to the digital era. 
It branches out from the base into 
two fundamental forms of money - tokens to the left and account claims to the right.
Examples are given in the era when they first appeared. 
Note some examples (such as shells and some crypto-assets) are not currently broadly
accepted as money.
\label{fig:money-tree}}}}
\end{center}
\end{figure}

While the history of money has been documented in considerable detail, we 
constructed a new summary visual representation to help us communicate
the evolution of digital money tokens.
This covers fours eras: the ancient and post-ancient eras (3000 BCE - 1300),
the modern era (1300 - 1950), and the digital era (1950 - present).
This is presented as the `money tree' in 
Figure~\ref{fig:money-tree} with the hope that others may also find it useful.

\subsection{Introduction to Money}
\label{sec:evolution-money-tree-money}

Money can be defined by the functions it
serves in society \cite{wealthnations-book}, which \cite{boe-speech-futuremoney}
describes as:

\begin{itemize}

  \item a \emph{store of value} with which to transfer purchasing power from today to
  some future time;
  
  \item a \emph{medium of exchange} with which to make payments for goods and services;
  and
  
  \item a \emph{unit of account} with which to measure the value of a particular good,
  service, saving or loan.
  
\end{itemize}

\noindent
As money is a social convention \cite{boe-speech-futuremoney}, performing the above
functions is not sufficient for something to be considered money because it also
needs to be readily and easily accepted by others.

The money tree branches out from the base into two fundamental forms of money: 
tokens\footnote{We use the term `token' to refer to all forms
of money that can be transferred directly from one party to another without relying on
an intermediary.
However, note there is currently no commonly agreed definition for this term, e.g.
some do not consider commodity money to be a token
\cite{mitchell-innes-money-paper},
some consider the term `object' to be more suitable than `token' \cite{imf-digital-money-tree},
and some consider crypto-assets to be an account-based form of money \cite{milne-bitcoin-account}.}
and account claims, or more simply, accounts \cite{mitchell-innes-money-paper}. 
Token-based money (e.g. cash) can be transferred directly from one party to another
without relying on an intermediary,
but account-based money (e.g. money in a bank account) 
cannot be transferred directly from one party to another and instead depends on
intermediaries (e.g. the bank(s) where the accounts are held) to effect transfers.
Another key distinction between token-based money and account-based money is the form of
verification needed when it is transferred, as described below (adapted
from \cite{bis-cbdcs}):

\begin{itemize}

  \item Token-based money (or payment systems) rely critically on the ability of the
  payee to verify the validity of the payment object. With cash a key risk is
  counterfeiting, while in the digital world the key risk is whether the token or `coin'
  is genuine or not (electronic counterfeiting) and whether it has already been spent.

  \item By contrast, systems based on account money depend fundamentally on the ability
  to verify the identity of the account holders.
  A key concern is identity theft, which allows perpetrators to transfer or withdraw
  money from accounts without permission. Identification
  is needed to correctly link payers and payees and to ascertain their respective
  account histories.

\end{itemize}

\vspace{1mm}

\subsection{Evolution of Account-based Money}
\label{sec:evolution-money-tree-account}

Account-based money appears to pre-date token-based money, based on available
information on ancient civilisations \cite{mitchell-innes-money-paper, moneyevolution-book}.
The evolution of account-based money is tied closely to the evolution of account
holding institutions. 

The `Account' side of the money tree depicts the following developments:

\begin{itemize}

  \item \emph{Ancient and Post-Ancient Eras:} Accounts in the ancient and post-ancient
  eras were held at trusted institutions,
  such as temples in Mesopotamia and Greece, where people could deposit precious
  metals or record credits in a currency. These deposits and credits were recorded on,
  for example, clay and marble tablets \cite{moneyevolution-book, banks-ages-book}.

  \item \emph{Modern Era:} Depositories developed into dedicated financial institutions
  during the Italian Renaissance, leading to the creation of official state banks.
  Account holding institutions evolved into three main types over the course of the modern
  era, partly in response to the growth of currencies: central banks, regulated financial
  institutions, and unregulated or non-financial institutions such as goldsmiths and early
  merchant banks \cite{banks-ages-book}.
  Each of these entities would hold accounts, usually denominated in a currency, with
  differing levels of access and trust. Transactions were recorded on paper ledgers using the
  `double-entry bookkeeping' system.
  While currencies were traditionally backed by commodity assets such as gold, towards
  the end of the modern era this relationship was removed with the introduction of `fiat 
  money' i.e. money that is not convertible to gold or any other asset \cite{boe-whats-money}.

  \item \emph{Digital Era:} The digital era has seen the development of digitised accounts
  that are well understood and widely deployed. Transactions continue to be recorded using
  the `double-entry bookkeeping' system using digital technologies. These have enabled
  account holders to have greater access to their money through digital and physical
  channels \cite{boe-speech-futuremoney}.
  While central banks and regulated financial institutions have digitised accounts and
  introduced new digital products (e.g. cards), the unregulated or non-financial
  institution space has also seen a rapid evolution of new digital products (e.g. mobile money,
  stored value digital wallets) from recent entrants such as telecommunication and technology
  firms.

\end{itemize}

\vspace{4mm}

\subsection{Evolution of Token-based Money}
\label{sec:evolution-money-tree-token}

The `Token' side of the money tree branches into two forms that have existed
from the ancient era to the digital era: `tokenised claims' that represent a claim on an
entity or right on an underlying asset (e.g. currency notes) and `claimless
tokens' that don't represent any such claim or right (e.g. coins). 
For tokenised claims, note the nature of the entity or asset on which the claim
or right lies can be an important factor although, for simplicity, this has not been
elaborated in the money tree.

The `Token' side of the money tree depicts the following developments:

\begin{itemize}

  \item \emph{Ancient and Post-Ancient Eras:} In the ancient and post-ancient eras,
  claimless tokens took the form
  of shells, beads or metal coins. 
  Tokenised claims, such as paper promissory notes and currency notes issued
  in ancient China, represented a claim on the issuing merchant or monarchy \cite{moneyevolution-book}.

  \item \emph{Modern Era:} The growth of currencies led to modern
  forms of money tokens created by dedicated financial institutions, non-financial
  institutions, central banks and governments.
  These entities issued metal coins and notes, representing the modern forms of claimless
  tokens and tokens that represented a claim or right \cite{wealthnations-book}.
  Some tokens represented a claim on account-based forms of money (e.g.
  negotiable instruments drawn on banks).
  Maintaining the value of the currency against the value of the metal coins
  and honouring the claims on notes issued were challenges for the issuers.
  This was due to variations in the subjective value of the metals in the coins
  and poor management of the relationship between notes issued and the underlying assets
  which backed them \cite{boe-speech-futuremoney}.
  With the introduction of fiat money towards the end of the modern era, the nature of
  the entity issuing the tokens became vital to determining a token's
  ability to serve the three functions of money (i.e. store of value,
  medium of exchange and unit of account).

  \item \emph{Digital Era:} The digital era has seen the birth of new digital forms of 
  money tokens \cite{boe-speech-futuremoney}.
  Similar to tokens in previous eras, digital money tokens can either be tokenised claims (e.g. central bank
  digital tokens \cite{bis-cbdcs}) or claimless tokens (e.g. some `cryptocurrencies').
  Again, the nature of the issuing entity remains critical to determining the
  token's ability to serve the three functions of money.

\end{itemize}

\vspace{5mm}

\section{Key Features of Digital Money Tokens}
\label{sec:key-features}

The field of digital money tokens is evolving rapidly. 
The terminology used by financial institutions, central
banks, regulators, international associations and fintechs to describe forms of digital money tokens
is not always consistent. 
This can make it challenging to understand the key features that
distinguish different forms of digital money tokens being developed and to evaluate new
product propositions. 
The current absence of a generally agreed taxonomy may also lead to a fragmented
regulatory approach \cite{wfe-taxonomy} potentially creating inappropriate opportunities for
regulatory arbitrage \cite{iosco-stablecoin-report}.

The seminal work by Bech and Garratt \cite{moneyflower} (that builds on earlier 
work \cite{bis-dig-currencies-paper, cbs-cbdc}) discusses and visually 
represents a taxonomy for money in the form of a `money flower', which is described
in the next section.
It has since been heavily referenced (e.g \cite{bcu-epeso,
actuaries-cbdc}) and also adapted (e.g. \cite{bis-cbdcs, 
agau-money-taxonomy}), becoming an key contribution to the taxonomy
of money. However, this high-level taxonomy does not consider all of the key
features of digital money tokens and so does not fully distinguish between these
forms of money.

Adrian and Mancini-Griffoli presented a visual
taxonomy for money \cite{imf-digital-money-tree} (coincidentally also named
`money tree') based on four features: type (claim or object), value (redemption rate for
claims, unit of account for objects), the backstop for claims (government or private),
and technology used (centralised or decentralisaed). This taxonomy comprises five
means of payment: central bank money, crypto-currency, b-money (issued by banks),
e-money (offered by new private sector providers), and i-money (issued by private
investment funds).
However, their use of `claim' and `object'
as the two distinct and fundamental types of money leads to listing cash and central
bank digital currencies as examples of `object' money even though they are
claims on central banks.

The Token Taxonomy Framework (TTF) \cite{ttf} is a taxonomy for digital tokens developed
by the InterWork Alliance, an industry association with members from the technology and 
financial industries.
The TTF is based on features such as fungibility, unit of measure
(fractional, whole or singleton), value (intrinsic or referential), uniqueness and the
nature of the supply (fixed, capped-variable, gated or infinite) that are combined with
behaviours (e.g. transferable, mintable, divisible) to create a set of base token types. 
These features, behaviours and base token types are generic and apply to
money tokens, securities tokens and utility tokens. 
This taxonomy does not appear to include all of the
key features of digital money tokens and is more focused on token implementation
mechanics.

Efforts by regulators and international associations to develop taxonomies for digital tokens
began in response to increasing consumer adoption of `cryptocurrencies'. 
This included a Financial Action Task Force (FATF) report \cite{fatf-virtualcurrency-defn}
defining `virtual currencies' and presenting a taxonomy based on convertibility and
centralisation.
These efforts to develop taxonomies continued partly in response to an increasing
number of Initial Coin Offerings (ICOs).
For example, 
guidelines by the Swiss Financial Market Supervisory Authority (FINMA) \cite{swiss-finma-guidelines}
used the term `payment token' and advice from the European Banking Authority (EBA) to
the European Commission  \cite{eba-advice-cryptoassets}
used the term `payment/exchange/currency token'.
The UK's Financial Conduct Authority (FCA) took a different approach in
its final guidance on crypto-assets \cite{fca-cryptoa-guidance} by defining a
taxonomy for tokens based on whether they are regulated. 
The two main categories were
`regulated tokens' (including `security tokens' and `e-money tokens') and
`unregulated tokens' (including `exchange tokens' and `utility tokens').

Recent consultations by BCBS \cite{bcbs-cryptoasset-consultation}, the European Commission
\cite{ecconsultation} and the FSB \cite{fsb-stablecoin-consultation}, revised guidance
from FINMA \cite{swiss-finma-newguidelines} and reports from BIS' Committee on Payments and
Market Infrastructures (CPMI) \cite{bis-stablecoin-paper} and the International Organization
of Securities Commissions (IOSCO) \cite{iosco-stablecoin-report} in response to the rise of
`stablecoins' provided a more detailed view of features of digital money tokens.
These features included stabilisation mechanisms, nature of
asset linkage, type of asset linked, nature of the claim, accessibility, reach,
redemption mechanics, permissioning model, transaction record model, and dependence on
existing payments infrastructure.
The IOSCO report states that ``\ldots many so-called stablecoins are neither
`stable' nor `coins' in the true sense of either word. So, whilst stablecoin is a
marketing term that has been widely adopted by industry, more neutral terms, may be more
accurate starting points for regulatory analysis''.

The European Central Bank (ECB) has defined a visual taxonomy for stablecoins \cite{ecb-stablecoin-taxonomy}
in the form of a `crypto-cube' based on three features: the existence of an
issuer responsible for satisfying any attached claim, whether decision-making
responsibilities over the stablecoin initiative are centralised, and what underpins the
value of a stablecoin (currency, other off-chain assets, on-chain assets or 
expectations). 
This taxonomy comprises four types of stablecoins: tokenised funds, off-chain
collateralised stablecoins, on-chain collateralised stablecoins and algorithmic
stablecoins.
Although the `crypto-cube' is a useful visual tool for categorising
stablecoins, it does not include all of the key features of digital money tokens.

GFMA's initial approach for a classification of crypto-assets \cite{gfma-fsb-response}
is based on four features: issuer,
mechanism or structure underlying the asset value (e.g. pegged to an underlying asset
or access to a service),
rights conferred (e.g. cash flows, redemption), and nature of the claim (e.g. claim on
issuer or claim on underlying asset). 
They propose a taxonomy comprising six types
of crypto-assets: `cryptocurrencies', `value-stable crypto-assets', `security tokens',
`settlement tokens', `utility tokens', and `other crypto-assets'.

\vspace{7mm}

\subsection{The Money Flower}
\label{sec:key-features-money-flower}

\begin{figure}[h!]
\begin{center}
\includegraphics[height=0.4\paperheight]{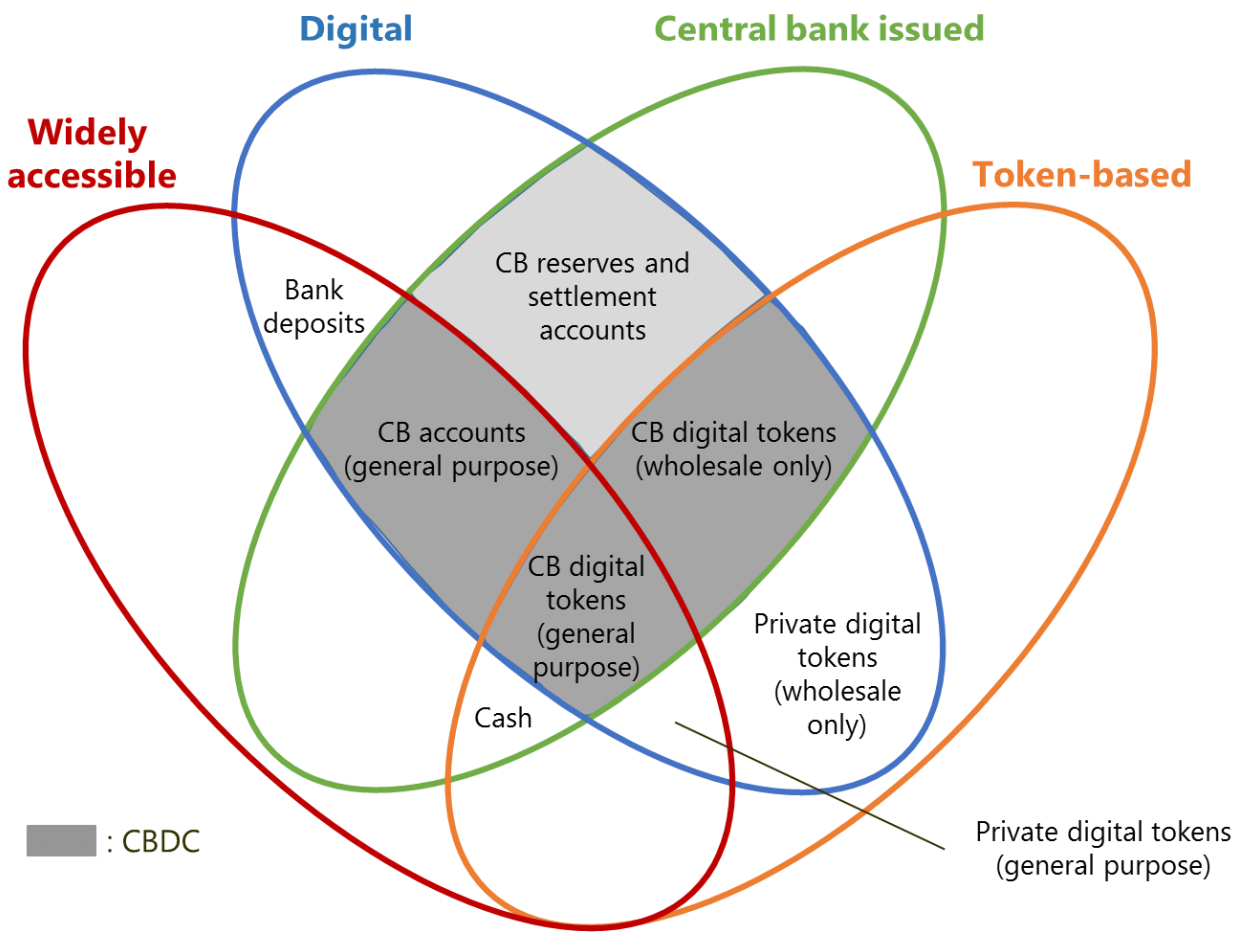}

\vspace{5mm}

\parbox{15cm}{\break\caption{\footnotesize {The Venn-diagram illustrates the four key
  properties of money.
  CB = central bank, CBDC = central bank digital currency (excluding digital central bank
  money already available to monetary counterparties and some non-monetary counterparties).
  Private digital tokens (general purpose) include crypto-assets and currencies, such as
  bitcoin and ethereum. 
  Source: Central Bank Digital Currencies, CPMI Papers. \textcopyright Bank for
  International Settlements 2018.
\label{fig:cbdc-money-flower}}}}
\end{center}
\end{figure}

Bech and Garratt created the money flower \cite{moneyflower} to support a new taxonomy
of money by showing how forms and specific examples of money 
fit into the overall monetary landscape.
In this section, we reproduce CPMI's adaption of the money flower \cite{bis-cbdcs}
in Figure~\ref{fig:cbdc-money-flower}
which depicts a taxonomy based on four features: 
issuer (central bank or not), form (digital or physical), accessibility (widely or
restricted), and technology (account-based or token-based).

CPMI describes other properties of money that are not depicted on their money
flower, including 24x7 availability, anonymity, whether
peer to peer transfers are supported, whether interest is borne, and
if any limits are imposed on holdings.

Digital money tokens are represented on CPMI's money flower as the
intersection of the `Digital' and `Token-based' ellipses, i.e. the four
areas labelled `CB digital tokens (wholesale only)',
`CB digital tokens (general purpose)',
`Private digital tokens (wholesale only)', and 
`Private digital tokens (general purpose)'.

\pagebreak

\subsection{Identifying the Key Features}
\label{sec:key-features-digital-money-tokens}

Mark Carney's speech on `The Future of Money' \cite{boe-speech-futuremoney}
explored the ability of `cryptocurrencies' 
to serve the three functions of money, i.e. store of value,
medium of exchange and unit of account.
In this section, we adopt a similar approach in order to evaluate the significance
of the features of digital money tokens and identify the key features
as follows:

\pagebreak

\begin{itemize}

  \item \emph{Issuer:} 
  An identifiable entity that controls the issuance of tokens, even if issued via a
  decentralised `smart contract' \footnote{We use the term `smart contract' to refer to
  ``\emph{\ldots an automatable and
  enforceable agreement. Automatable by computer, although some parts may require human
  input and control. Enforceable either by legal enforcement of rights and obligations or
  via tamper-proof execution of computer code}'' \cite{clack2016smart}} or an approved
  intermediary. 
  While some digital money token arrangements do not have a specific issuer,
  in other arrangements an identifiable entity exercises
  control over this process even if token issuance is handled by a decentralised
  smart contract or an approved intermediary. 
  The nature of this entity, its decisions regarding other key features (nature of claim,
  asset linkage, type of linked asset, redemption etc) and its ability to operate the
  token arrangement appropriately determine whether the digital money tokens issued can
  fulfill all three functions of money. 
  As a result, this is likely to be the most significant feature.
  
  \item \emph{Claim, Right or Interest:}
  The nature of the claim, right or interest arising from ownership or control of a 
  token.
  Depending on the digital token arrangement, the ownership of digital money tokens
  may or may not confer the owner with a claim, right or interest.
  The nature of this claim, right or interest can influence the 
  token's ability to serve as a store of value.
  
  \item \emph{Asset Linkage:}
  The nature of the linkage, if any, between the value of the token and an underlying
  asset or basket of assets.
  The linkage can be to part or all of an underlying asset.
  The nature of this linkage can affect the stability of the digital money tokens'
  value and can thereby influence the token's ability to serve as a store of value.

  \item \emph{Type of Linked Asset:}
  The type of asset linked to the token.
  The type of the linked asset can influence the token's ability to serve as a store of
  value because volatility in the value of the asset will affect the value of the tokens.
  In addition, some linked asset types (e.g. cash denominated in a fiat currency) can
  also naturally serve as an underlying unit of account.
  
  \item \emph{Redemption Rate \footnote{We use the term `redemption rate' instead of
  `exchange rate' \cite{bis-stablecoin-paper} 
  because `exchanging' the token is a more general term that includes other actions
  such as buying 
  other assets from any entity using a token}:}
  The rate at which tokens are redeemed by the issuer or an approved intermediary against
  a pre-determined asset or set of assets.
  If the digital money token arrangement supports redemption of
  issued tokens, whether the redemption rate is fixed or variable impacts the stability
  of the token's value, which can influence the token's ability to serve as a store
  of value. 
  
  \item \emph{Denomination:}
  The unit of account used to quantify token ownership.
  The denomination of a digital money token can affect its ability to
  serve as a unit of account. 
  For example,
  retailers that quote prices in `cryptocurrencies' typically update them 
  frequently due to their volatility.
  
  \item \emph{Accessibility:}
  Accessibility distinguishes between tokens that are available everywhere to everyone
  and tokens that are restricted to certain agents or uses.
  This feature refers to restrictions, typically imposed by 
  the issuer, such as only specific types of entities that can participate 
  (e.g. regulated financial institutions) or only specific uses (e.g. 
  wholesale banking transactions).
  Accessibility can influence the extent to which digital money tokens can 
  serve as a medium of exchange.
  
\end{itemize}

\vspace{5mm}

\noindent
As discussed previously, money is a social convention and it is therefore not sufficient
for a digital money token to have all the above key features to be considered money because
it also needs to be readily and easily accepted.

In addition to these key features, there are other features such as  
availability, anonymity, whether interest is borne, holding limits, reach, permissioning
model, transaction record model, nature of supply, and dependence on existing payments
infrastructure.
While these features represent important design choices, they are likely to be of
lesser significance because they are either subsumed under the above key features or have a
smaller impact on whether a given digital money token can serve the three 
functions of money.

\vspace{5mm}

\subsection{Options for the Key Features}
\label{sec:feature-options}

\begin{figure}[h!]
\begin{center}
\includegraphics[height=0.27\paperheight]{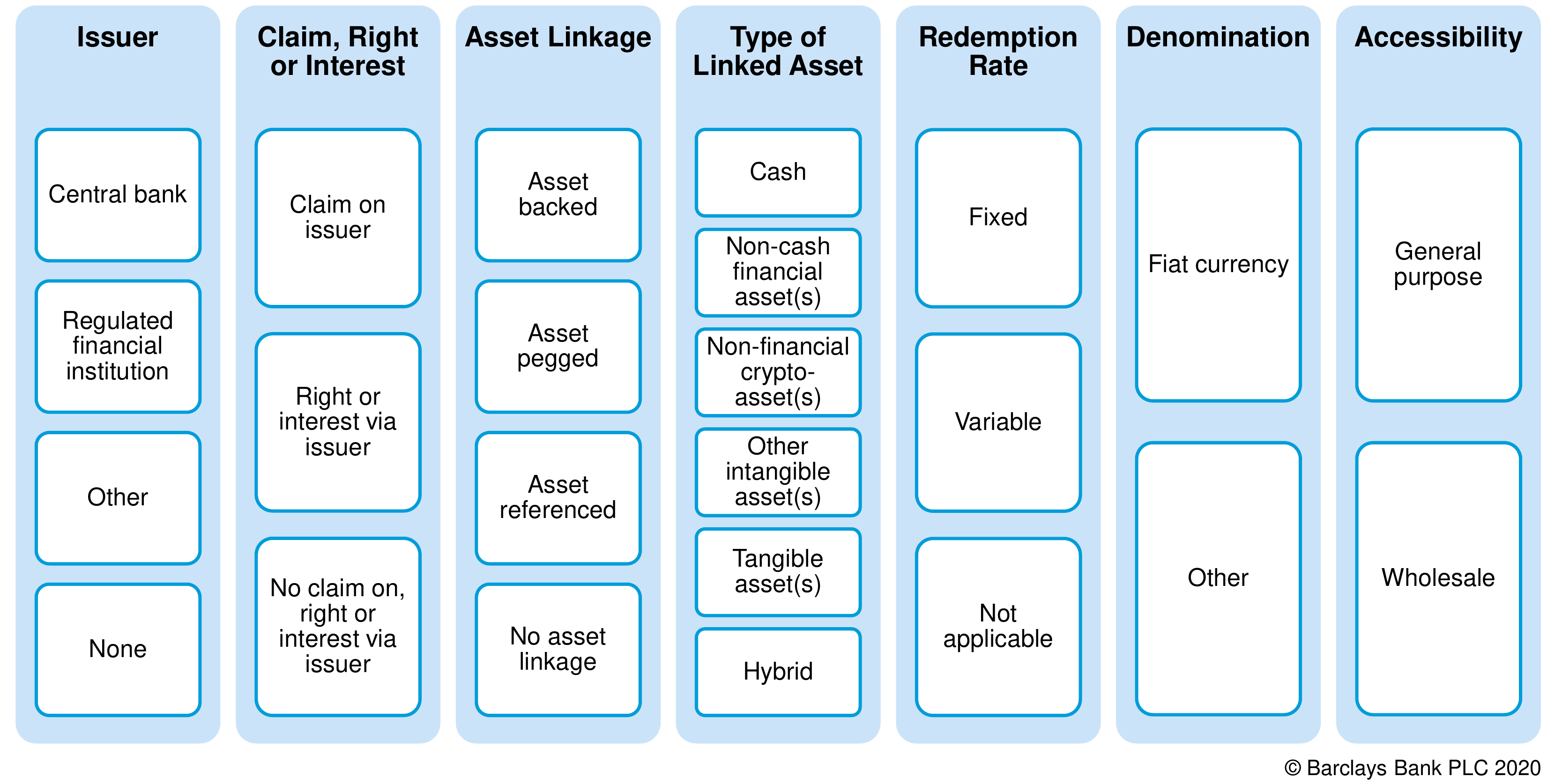}

\vspace{3mm}

\parbox{15cm}{\caption{\footnotesize {Summary of key features and options for digital money tokens
\label{fig:features-options}}}}
\end{center}
\end{figure}

In order to develop a robust taxonomy, a list of key features and options is
necessary.
We now examine the potential options for each of the key features of digital money tokens,
as summarised in Figure~\ref{fig:features-options}.

\noindent
Table \ref{table:digitaltokenmoney-features-options} below describes these options for key features
of digital money tokens, together with illustrative examples.

\pagebreak

\begin{center}
\begingroup
\renewcommand{\arraystretch}{1.5}     
\captionsetup{width=15cm}
  
  \begin{longtable}{>{\raggedright}p{0.22\textwidth} >{\raggedright}p{0.52\textwidth} >{\raggedright\arraybackslash}p{0.17\textwidth}}
   
   \toprule
   \textbf{Feature/Option} & \textbf{Description}  & \textbf{Example(s)}  \\  [0.5ex] 
   \midrule
   \endfirsthead

   \multicolumn{3}{r}{\textit{Continued from previous page}} \\
   \toprule
   \textbf{Feature/Option} & \textbf{Description}  & \textbf{Example(s)}  \\  [0.5ex] 
   \midrule
   \endhead

   \bottomrule
   \multicolumn{3}{r}{\textit{Continued on next page}} \\
   \endfoot

   \bottomrule
   \caption{\footnotesize{The options for each of the key features of digital money
   tokens, with brief descriptions and illustrative examples.}}
   \label{table:digitaltokenmoney-features-options}
   \endlastfoot

    \multicolumn{3}{l}{\emph{Issuer}} \\ 
    
    Central bank & Financial institution with privileged control over the production and 
    distribution of money &
    e-Krona {\footnotesize\cite{ekrona}} \\ 
    
    Regulated Financial Institution (FI) & Banks, Financial Market Infrastructures
    (FMIs) and other regulated FIs & USC {\footnotesize\cite{fnality}}, \hspace{5em}
    JPM Coin {\footnotesize\cite{jpmcoin}} \\ 

    Other & Unregulated or non-financial entity & Tether {\footnotesize\cite{tether-paper}} \\ 

    None & No specific entity controls the issuance of tokens, issuance is managed entirely
    algorithmically & Bitcoin \\ 

   \midrule

    \multicolumn{3}{l}{\emph{Claim, Right or Interest}} \\ 
    
    Claim on issuer & Legal claim on the issuer or an approved intermediary, typically
    exercised via instruction to the issuer or an approved intermediary & e-Krona, \hspace{5em}
    JPM Coin \\

    Right or interest via issuer & No legal claim on the issuer, but right or interest
    (e.g. share of underlying asset) exercised via instruction to the issuer or
    an approved intermediary e.g. to redeem share & USC \\ 

    No claim on, right or interest via issuer & No legal claim on the issuer, no right or
    interest exercised via issuer or no specific issuer & Bitcoin \\

   \midrule

    \multicolumn{3}{l}{\emph{Asset Linkage}} \\ 
   
    Asset backed & Tokens are fully backed by underlying assets that are typically held or
    controlled by the issuer or an approved intermediary & USC, Single-currency Libra Coin {\footnotesize\cite{libra}}  \\

    Asset pegged & Tokens are not fully backed by assets but their value is fixed against
    one or more assets, typically by leveraging the financial strength and stability of
    the issuer & - \\

    Asset referenced & Tokens are not fully backed by or pegged to assets but refer to
    assets to determine value & Dai {\footnotesize\cite{makerdao-whitepaper}} \\

    No asset linkage & Tokens are not fully backed by, pegged to or referenced to any
    assets & Bitcoin \\

   \pagebreak

    \multicolumn{3}{l}{\emph{Type of Linked Asset}
    \footnote{We use the terms `tangible assets', `cash', `financial assets' and
    `intangible assets' based on the International Financial Reporting
    Standards (IFRS). The treatment of `crypto-assets' is based on guidance from the
    IFRS Interpretations Committee \cite{ifric-cryptocurrencies} and the Institute of
    Singapore Chartered Accountants \cite{isca-crypoassets}.
    Linkage to an asset is different from linkage
    to a basket of assets of the same type, but these options are listed together
    for brevity}} \\ 

    Cash & Tokens are linked to cash denominated in a fiat currency & USC \\

    Non-cash financial asset(s) & Tokens are linked to a non-cash financial asset or a
    basket of such financial assets (e.g. securities, contractual rights) & 
    ArCoin {\footnotesize\cite{forbes-arcoin}} \\

    Non-financial crypto-asset(s) & Tokens are linked to a crypto-asset or a basket of
    crypto-assets that do not represent cash or other financial assets
    (e.g. Ether) &  Dai \\

    Other intangible asset(s) & Tokens are linked to an intangible asset not covered by
    the above options (e.g. intellectual property, digital goods) or a basket of such
    assets & - \\

    Tangible asset(s) & Tokens are linked to a tangible asset (e.g commodities, real estate) or
    a basket of tangible assets & PMGT {\footnotesize\cite{pmgt-whitepaper}} \\

    Hybrid & Tokens are linked to a basket of two or more of the above asset types &
    Single-currency Libra Coin \\

   \midrule

    \multicolumn{3}{l}{\emph{Redemption Rate}} \\ 

    Fixed & Tokens can be redeemed against
    pre-determined assets at a fixed rate & USC, \hspace{5em} JPM Coin \\

    Variable & Tokens can be redeemed against
    pre-determined assets at a variable rate & Multi-currency Libra Coin \\

    Not applicable & Tokens cannot be redeemed or there is no issuer & Bitcoin \\

   \midrule

    \multicolumn{3}{l}{\emph{Denomination}} \\ 

    Fiat currency & Tokens are denominated in a fiat currency or are redeemable in
    a fiat currency at a fixed rate & e-Krona, \hspace{5em} USC \\

    Other & Tokens have their own independent denomination & Multi-currency
    Libra Coin \\

   \pagebreak

    \multicolumn{3}{l}{\emph{Accessibility}} \\ 

    General purpose & Tokens are intended for general use, including both retail and
    wholesale purposes & e-Krona, \hspace{5em} Libra Coin \\

    Wholesale & Tokens where access is restricted, for example to FIs or selected 
    clients of FIs & USC, \hspace{5em} JPM Coin \\ [1ex] 

  \end{longtable}

\endgroup

\end{center}

\section{Summary}
\label{sec:conclusion}

In this paper, we presented a `money tree' which placed the recent concept of
digital money tokens into a historical context by illustrating their evolution from
more traditional forms of money.
We then identified the key features of digital money tokens with options
and examples.
The money tree and the list of key features and options contributes
to the design space for the emerging field of digital money tokens.
We hope this paper will be of interest to the financial services
industry as it innovates with digital money tokens.
We welcome feedback and look forward to continuing industry collaboration on the
the classification of digital money tokens. 
In particular, international collaboration between financial
institutions, central banks, regulators, associations and fintechs
will enable the development of global standards.

\section*{Acknowledgements}
\label{sec:acknowledgements}

The author would like to thank Lee Braine (Barclays) for direction, input and review
of this paper. 
Thanks are also due to Nicole Sandler (Barclays), Vikram Bakshi (Barclays), Richard
Barnes (Barclays) and Simon Gleeson (Clifford Chance) for their helpful feedback.

\noindent

\pretolerance=-1
\tolerance=-1
\emergencystretch=0pt

\bibliography{digital-money-tokens}

\begin{thebibliography}{10}

\bibitem{imf-digital-money-tree}
Tobias Adrian and Tommaso Mancini-Griffoli.
\newblock {The Rise of Digital Money}.
\newblock {FinTech Note} 19/01, {International Monetary Fund}, 2019.
\newblock
  \url{https://www.imf.org/~/media/Files/Publications/FTN063/2019/English/FTNEA2019001.ashx}.

\bibitem{afme-ec-response}
{Association for Financial Markets in Europe}.
\newblock {Consultation response: European Commission Public Consultation - An
  EU Framework for Markets in Crypto-assets}.
\newblock 2020.
\newblock
  \url{https://www.afme.eu/Portals/0/DispatchFeaturedImages/2020%2003%2019%20AFME%20EC%20Legal%20Framework%20for%20Crypto-assets.pdf}.

\bibitem{boe-whats-money}
{Bank of England}.
\newblock {What is money?}
\newblock \url{https://www.bankofengland.co.uk/knowledgebank/what-is-money},
  2020.
\newblock Accessed 13 July 2020.

\bibitem{bcbs-cryptoasset-consultation}
{Basel Committee on Banking Supervision}.
\newblock {Designing a prudential treatment for cryptoassets}.
\newblock Discussion paper, {BIS}, 2019.
\newblock \url{https://www.bis.org/bcbs/publ/d490.pdf}.

\bibitem{moneyflower}
Morten Bech and Rodney Garratt.
\newblock {Central bank cryptocurrencies}.
\newblock In {\em {BIS Quarterly Review}}, pages 55--70. BIS, September 2017.
\newblock \url{https://www.bis.org/publ/qtrpdf/r\_qt1709f.pdf}.

\bibitem{cbs-cbdc}
Ole Bjerg.
\newblock {Designing New Money: The Policy Trilemma of Central Bank Digital
  Currency}.
\newblock {Working Paper}, {Copenhagen Business School}, 2017.
\newblock
  \url{https://research-api.cbs.dk/ws/portalfiles/portal/58550948/Designing\_New\_Money\_The\_policy\_trilemma\_of\_central\_bank\_digital\_currency.pdf}.

\bibitem{forbes-arcoin}
Charles Bovaird.
\newblock {Arca Uses Ethereum In First SEC-Registered Fund For Digital Shares}.
\newblock
  \url{https://www.forbes.com/sites/cbovaird/2020/07/06/arca-launches-first-sec-registered-fund-to-offer-digital-shares},
  2020.
\newblock Accessed 20 July 2020.

\bibitem{boe-speech-futuremoney}
Mark Carney.
\newblock {The Future of Money}.
\newblock In {\em Scottish Economics Conference}. Edinburgh University, 2018.
\newblock
  \url{https://www.bankofengland.co.uk/-/media/boe/files/speech/2018/the-future-of-money-speech-by-mark-carney.pdf}.

\bibitem{clack2016smart}
Christopher~D Clack, Vikram~A Bakshi, and Lee Braine.
\newblock Smart contract templates: foundations, design landscape and research
  directions.
\newblock {\em The Computing Research Repository (CoRR)}, abs/1608.00771, 2016.
\newblock \url{https://arxiv.org/pdf/1608.00771}.

\bibitem{bis-dig-currencies-paper}
{Committee on Payments and Market Infrastructures}.
\newblock {Digital Currencies}.
\newblock {CPMI Papers} 137, {BIS}, 2015.
\newblock \url{https://www.bis.org/cpmi/publ/d137.pdf}.

\bibitem{bis-cbdcs}
{Committee on Payments and Market Infrastructures}.
\newblock {Central bank digital currencies}.
\newblock {CPMI Papers} 174, {BIS}, 2018.
\newblock \url{https://www.bis.org/cpmi/publ/d174.pdf}.

\bibitem{ecconsultation}
{Directorate-General for Financial Stability, Financial Services and Capital
  Markets Union}.
\newblock {On an EU framework for markets in crypto-assets}.
\newblock Consultation document, {European Commission}, 2019.
\newblock
  \url{https://ec.europa.eu/info/sites/info/files/business\_economy\_euro/banking\_and\_finance/documents/2019-crypto-assets-consultation-document\_en.pdf}.

\bibitem{eba-advice-cryptoassets}
{European Banking Authority}.
\newblock {Report with advice for the European Commission on crypto-assets}.
\newblock 2019.
\newblock
  \url{https://eba.europa.eu/sites/default/documents/files/documents/10180/2545547/67493daa-85a8-4429-aa91-e9a5ed880684/EBA%20Report%20on%20crypto%20assets.pdf}.

\bibitem{ecb-stablecoin-taxonomy}
{European Central Bank}.
\newblock {Stablecoins - no coins, but are they stable?}
\newblock {In Focus,} Issue No 3, 2019.
\newblock
  \url{https://www.ecb.europa.eu/paym/intro/publications/pdf/ecb.mipinfocus191128.en.pdf}.

\bibitem{fatf-virtualcurrency-defn}
{Financial Action Task Force}.
\newblock {Virtual Currencies: Key Definitions and Potential AML/CFT Risks}.
\newblock 2014.
\newblock
  \url{http://www.fatf-gafi.org/media/fatf/documents/reports/Virtual-currency-key-definitions-and-potential-aml-cft-risks.pdf}.

\bibitem{fca-cryptoa-guidance}
{Financial Conduct Authority}.
\newblock {Guidance on Cryptoassets: Feedback and Final Guidance to CP 19/3}.
\newblock Policy Statement PS19/22, 2019.
\newblock \url{https://www.fca.org.uk/publication/policy/ps19-22.pdf}.

\bibitem{fsb-stablecoin-consultation}
{Financial Stability Board}.
\newblock {Addressing the regulatory, supervisory and oversight challenges
  raised by ``global stablecoin'' arrangements}.
\newblock Consultative document, 2020.
\newblock \url{https://www.fsb.org/wp-content/uploads/P140420-1.pdf}.

\bibitem{fnality}
{Fnality International}.
\newblock {What we do}.
\newblock \url{https://www.fnality.org/what-we-do}, 2019.
\newblock Accessed 30 June 2020.

\bibitem{makerdao-whitepaper}
Maker Foundation.
\newblock {The Maker Protocol: MakerDAO's Multi-Collateral Dai (MCD) System}.
\newblock \url{https://makerdao.com/en/whitepaper}, 2019.
\newblock Accessed 29 July 2020.

\bibitem{bis-stablecoin-paper}
{G7 Working Group on Stablecoins}.
\newblock {Investigating the Impact of Global Stablecoins}.
\newblock {CPMI Papers} 187, {BIS}, 2019.
\newblock \url{https://www.bis.org/cpmi/publ/d187.pdf}.

\bibitem{gfma-bcbs-response}
{Global Financial Markets Association}.
\newblock {Consultation response: Basel Committee on Banking Supervision -
  Designing a Prudential Treatment for Crypto-Assets}.
\newblock 2020.
\newblock
  \url{https://www.gfma.org/wp-content/uploads/2020/04/gfma-bcbs-prudential-crypto-assets-final-consolidated-version-20200427.pdf}.

\bibitem{gfma-fsb-response}
{Global Financial Markets Association}.
\newblock {Consultation Response: Financial Stability Board - Addressing the
  regulatory, supervisory and oversight challenges raised by ``global
  stablecoin'' arrangements}.
\newblock 2020.
\newblock \url{https://www.fsb.org/wp-content/uploads/GFMA-1.pdf}.

\bibitem{banks-ages-book}
Noble~Foster Hoggson.
\newblock {\em {Banking Through the Ages}}.
\newblock Dodd, Mead and Company, New York, 1926.

\bibitem{ifric-cryptocurrencies}
{IFRS Interpretations Committee (IFRIC)}.
\newblock {Holding of Cryptocurrencies}.
\newblock {IFRIC Interpretations}, {International Accounting Standards Board},
  June 2019.
\newblock
  \url{https://cdn.ifrs.org/-/media/feature/supporting-implementation/agenda-decisions/holdings-of-cryptocurrencies-june-2019.pdf}.

\bibitem{pmgt-whitepaper}
InfiniGold.
\newblock {Perth Mint Gold Token: Whitepaper ver. 1.1}.
\newblock \url{https://pmgt.io/static/assets/pmgt\_whitepaper.pdf}, 2019.
\newblock Accessed 20 July 2020.

\bibitem{isca-crypoassets}
{Institute of Singapore Chartered Accountants}.
\newblock {Accounting for Cryptoassets: From a Holder's Perspective}.
\newblock {ISCA Financial Reporting Guidance} FRG 2, 2020.
\newblock
  \url{https://isca.org.sg/media/2824062/frg-2-accounting-for-cryptoassets-from-a-holder-s-perspective.pdf}.

\bibitem{ttf}
{InterWork Alliance}.
\newblock {Token Taxonomy Framework}.
\newblock \url{https://github.com/InterWorkAlliance/TokenTaxonomyFramework},
  2020.
\newblock Accessed 07 July 2020.

\bibitem{uot-moneyevolution-paper}
Katsuhito Iwai.
\newblock {Evolution of Money}.
\newblock In Ugo Pagano and Antonio Nicita, editors, {\em Evolution of Economic
  Diversity}, pages 396--441. Routledge, 1997.
\newblock \url{https://ssrn.com/abstract=1861952}.

\bibitem{jpmcoin}
{J.P. Morgan}.
\newblock {J.P. Morgan Creates Digital Coin for Payments}.
\newblock \url{https://www.jpmorgan.com/global/news/digital-coin-payments},
  2019.
\newblock Accessed 30 June 2020.

\bibitem{libra}
{Libra Association Members}.
\newblock {Libra White Paper v2.0}.
\newblock \url{https://libra.org/en-US/white-paper}, 2020.
\newblock Accessed 30 June 2020.

\bibitem{bcu-epeso}
Gerardo Licandro.
\newblock {Uruguayan e-Peso on the context of financial inclusion}.
\newblock In {\em {Conference on ``Economics of Payments IX''}}. BIS, 2018.
\newblock \url{https://www.bis.org/events/eopix\_1810/licandro\_pres.pdf}.

\bibitem{milne-bitcoin-account}
Alistair Milne.
\newblock {Argument by False Analogy: The Mistaken Classification of Bitcoin as
  Token Money}.
\newblock 2018.
\newblock \url{https://papers.ssrn.com/sol3/papers.cfm?abstract\_id=3290325}.

\bibitem{mitchell-innes-money-paper}
Alfred Mitchell-Innes.
\newblock {What is money?}
\newblock {\em The Banking Law Journal}, pages 377--408, 1913.
\newblock
  \url{https://www.newmoneyhub.com/www/money/mitchell-innes/what-is-money.html}.
  Accessed 30 June 2020.

\bibitem{mas-ubin-5}
{Monetary Authority of Singapore and Temasek}.
\newblock {Project Ubin Phase 5: Enabling Broad Ecosystem Opportunities}.
\newblock 2020.
\newblock
  \url{https://www.mas.gov.sg/-/media/MAS/ProjectUbin/Project-Ubin-Phase-5-Enabling-Broad-Ecosystem-Opportunities.pdf}.

\bibitem{moneyevolution-book}
David Orrell and Roman Chlupatý.
\newblock {\em {The Evolution of Money}}.
\newblock Columbia University Press, 2016.

\bibitem{agau-money-taxonomy}
Joe Thierry~Arys Ruiz and Abdul-Hadi~Bashir Subhia.
\newblock Agau and the new taxonomy of money.
\newblock {Research Report}, Tarco International, 2018.
\newblock
  \url{https://www.researchgate.net/profile/Joe\_Thierry\_Ruiz/publication/330564376\_AGAU\_AND\_THE\_NEW\_TAXONOMY\_OF\_MONEY/links/5d6414d5299bf1f70b0ea813/AGAU-AND-THE-NEW-TAXONOMY-OF-MONEY.pdf}.
  Accessed 07 July 2020.

\bibitem{wealthnations-book}
Adam Smith.
\newblock {\em {An Inquiry into the Nature and Causes of the Wealth of
  Nations}}.
\newblock {W. Strahan and T. Cadell}, London, 1776.

\bibitem{ekrona}
{Sveriges Riksbank}.
\newblock {E-krona}.
\newblock \url{https://www.riksbank.se/en-gb/payments--cash/e-krona}, 2019.
\newblock Accessed 30 June 2020.

\bibitem{swiss-finma-guidelines}
{Swiss Financial Market Supervisory Authority}.
\newblock {Guidelines for enquiries regarding the regulatory framework for
  initial coin offerings (ICOs)}.
\newblock 2018.
\newblock
  \url{https://www.finma.ch/en/~/media/finma/dokumente/dokumentencenter/myfinma/1bewilligung/fintech/wegleitung-ico.pdf?la=en}.

\bibitem{swiss-finma-newguidelines}
{Swiss Financial Market Supervisory Authority}.
\newblock {Supplement to the guidelines for enquiries regarding the regulatory
  framework for initial coin offerings (ICOs)}.
\newblock 2019.
\newblock
  \url{https://www.finma.ch/en/~/media/finma/dokumente/dokumentencenter/myfinma/1bewilligung/fintech/wegleitung-stable-coins.pdf?la=en}.

\bibitem{tether-paper}
{Tether}.
\newblock {Tether: Fiat currencies on the Bitcoin blockchain}.
\newblock
  \url{https://tether.to/wp-content/uploads/2016/06/TetherWhitePaper.pdf},
  2016.
\newblock Accessed 21 August 2020.

\bibitem{iosco-stablecoin-report}
{The Board of the International Organization of Securities Commissions}.
\newblock {Global Stablecoin Initiatives}.
\newblock {Public Report} OR01/2020, {IOSCO}, 2020.
\newblock \url{https://www.iosco.org/library/pubdocs/pdf/IOSCOPD650.pdf}.

\bibitem{actuaries-cbdc}
Orla Ward and Sabrina Rochemont.
\newblock {Understanding Central Bank Digital Currencies (CBDC)}.
\newblock {An addendum to ``A Cashless Society - Benefits, Risks and Issues
  (Interim paper)''}, {The Institute and Faculty of Actuaries}, 2019.
\newblock
  \url{https://www.actuaries.org.uk/system/files/field/document/Understanding%20CBDCs%20Final%20-%20disc.pdf}.

\bibitem{wfe-taxonomy}
{World Federation of Exchanges}.
\newblock {WFE Response to the Financial Stability Board's Consultation
  Document}.
\newblock 2020.
\newblock
  \url{https://www.world-exchanges.org/storage/app/media/wfe-response-to-the-fsbs-consultation-on-global-stablecoin.pdf}.

\end{thebibliography}
\bibliographystyle{plain}

\end{document}